\documentclass{article}
\usepackage{frascatiphys}
\begin{document}
\title{THE STRANGENESS PHYSICS PROGRAM AT CLAS}
\author{Daniel S. Carman\\
{\em Jefferson Laboratory, 12000 Jefferson Ave., Newport News, VA, U.S.A.}\\
{\em carman@jlab.org}}
\maketitle
\baselineskip=11.6pt
\begin{abstract}
An extensive program of strange particle production off the nucleon is 
currently underway with the CEBAF Large Acceptance Spectrometer (CLAS) 
in Hall B at Jefferson Laboratory. This talk will emphasize strangeness 
electroproduction in the baryon resonance region between $W$=1.6 and 
2.4~GeV, where indications of $s$-channel structure are suggestive of 
high-mass baryon resonances coupling to kaons and hyperons in the final 
state.  Precision measurements of cross sections and polarization 
observables are being carried out with highly polarized electron and real 
photon beams at energies up to 6~GeV.  The near-term and longer-term
future of this program will also be discussed.
\end{abstract}
\baselineskip=14pt
\section{Introduction}

An important key to understand the structure of the nucleon is to understand
its spectrum of excited states.  However, understanding nucleon resonance
excitation provides a serious challenge to hadronic physics due to the
non-perturbative nature of QCD at these energies.  Recent symmetric quark 
model calculations predict more states than have been seen experimentally
\cite{capstick}.  Mapping out the spectrum of these excited states will 
provide for insight into the underlying degrees of freedom of the nucleon.

Most of our present knowledge of baryon resonances comes from reactions
involving pions in the initial and/or final states.  A possible explanation
for the so-called missing resonance problem could be that pionic coupling
to the intermediate $N^*$ or $\Delta^*$ states is weak.  This
suggests a search for these hadronic states in strangeness production
reactions.  Beyond different coupling constants (e.g. $g_{KNY}$ vs. 
$g_{\pi NN}$), the study of the exclusive production of $K^+\Lambda$ and 
$K^+\Sigma^0$ final states has other advantages in the search for missing 
resonances.  The higher masses of the kaon and hyperons, compared to their 
non-strange counterparts, kinematically favor a two-body decay mode for 
resonances with masses near 2~GeV, a situation that is experimentally 
advantageous.  In addition, baryon resonances have large widths and are often 
overlapping.  Studies of different final states can provide for important 
cross checks in quantitatively understanding the contributing amplitudes.  
Although the two ground-state hyperons have the same valence quark 
structure ($uds$), they differ in isospin, such that intermediate $N^*$ 
resonances can decay strongly to $K^+\Lambda$ final states, while both 
$N^*$ and $\Delta^*$ decays can couple to $K^+\Sigma^0$ final states.

The search for missing resonances requires more than identifying features 
in the mass spectrum. QCD cannot be directly tested with $N^*$ spectra 
without a model for the production dynamics\cite{satolee}. The $s$-channel 
contributions are known to be important in the resonance region in order to 
reproduce the invariant mass ($W$) spectra, while $t$-channel meson exchange 
is also necessary to describe the diffractive part of the production and 
$u$-channel diagrams are necessary to describe the back-angle strength.  
Thus measurements that can constrain the phenomenology for these reactions 
are just as important as finding one or more of the missing resonances.  

Theoretically, there has been considerable effort during the past decade to 
develop models for the $KY$ photo- and electroproduction processes.  However, 
the present state of understanding is still limited by a sparsity of data.  
Model fits to the existing cross section data are generally obtained at the 
expense of many free parameters, which leads to difficulties in constraining 
theory.  Moreover, cross section data alone are not sufficient to fully 
understand the reaction mechanism, as they probe only a portion of the 
full response.  In this regard, measurements of spin observables are essential 
for continued theoretical development in this field, as they allow for 
improved understanding of the dynamics of this process and provide for 
strong tests of QCD-inspired models.

In this talk I focus on the strangeness electroproduction program in Hall B 
at Jefferson Laboratory using the CLAS detector\cite{mecking}. Presently 
there is very limited knowledge of $N^*,\Delta^* \to KY$ couplings.  With 
the existing CLAS program, the present lack of data will be remedied with a 
wealth of high quality measurements spanning a broad kinematic range.

\section{Theoretical Models}
\label{sec:theory}

With the recently available data from the photo- and electroproduction of
$KY$ final states from CLAS and elsewhere, there have been renewed efforts
on the development of theoretical models.  The majority of these are
single-channel models that represent tree-level calculations, where the 
amplitude is constructed from the lowest-order Feynman diagrams.  More 
recent work has moved beyond the single-channel approach with the development 
of coupled-channels models\cite{juliadiaz,chiang,shklyar,leecole} or by 
fitting simultaneously to multiple but independent reaction channels
\cite{sarantsev,anisov}.  However, as a combined coupled-channels 
analysis of the photo- and electroproduction reactions is not yet available, 
a tree-level approach currently represents the best possibility of studying 
both reactions within the same framework.  While most of the recent 
theoretical analyses have focussed solely on the available photoproduction 
data, it has been shown that electroproduction observables can yield 
important complementary insights to improve and constrain the theory.

At JLab energies, perturbative QCD is not yet capable of providing 
analytical predictions for the observables for kaon electroproduction.  In 
order to understand the underlying physics, effective models must be employed 
that ultimately represent approximations to QCD.  In this work we compare 
our data against three different model approaches.  The first is a 
traditional hadrodynamic (resonance) model, the second is based on a 
Reggeon-exchange model, and the third is a hybrid Regge plus resonance 
approach.  

In the hadrodynamic model approach, the strong interaction is modeled by an 
effective Lagrangian, which is constructed from tree-level Born and extended 
Born terms for intermediate states exchanged in the $s$, $t$, and $u$ 
reaction channels.  Each resonance has its own strong coupling constants and 
strong decay widths.  A complete description of the physics processes requires 
taking into account all possible channels that could couple to the initial 
and final state measured, but the advantages of the tree-level approach 
include the ability to limit complexity and to identify the dominant trends.  
In the one-channel, tree-level approach, several dozen parameters must be 
fixed by fitting to the data, since they are poorly known and not constrained 
from other sources.  

The hadrodynamic models shown in this work were developed by Mart and 
Bennhold\cite{mart} and the Ghent group\cite{janssen}.  In 
these models, the coupling strengths have been determined mainly by fits 
to existing $\gamma p \to K^+ Y$ data (with some older electroproduction 
data included), leaving the coupling constants as free 
parameters (constrained loosely by SU(3) symmetry requirements).  The model 
parameters are not based on fits to any CLAS data.  The specific resonances 
included with these models include the $S_{11}$(1650), $P_{11}$(1710), 
$P_{13}$(1720), and $D_{13}$(1895) $N^*$ states in the $s$-channel, and the 
$K^*$(892) and $K^*_1$(1270) in the $t$-channel.  The Ghent model also
includes hyperon exchange in the $K^+\Lambda$ $u$-channel and couplings
of $s$-channel $S_{31}$(1900) and $P_{31}$(1910) $\Delta^*$ states for the
$K^+\Sigma^0$ final state.

In this work, we also compare our results to the Reggeon-exchange model 
from Guidal {\it et al.}\cite{guidal}.  This calculation includes no baryon 
resonance terms at all.  Instead, it is based only on gauge-invariant 
$t$-channel $K$ and $K^*$ Regge-trajectory exchange.  It therefore provides 
a complementary basis for studying the underlying dynamics of strangeness 
production.  It is important to note that the Regge approach has far fewer 
parameters compared to the hadrodynamic models.  These include the $K$ and 
$K^*$ form factors and the coupling constants $g_{KYN}$ and $g_{K^*YN}$ 
(taken from photoproduction studies).

The final model included in this work was also developed by the Ghent 
group\cite{ghent}, and is based on a tree-level effective field model for 
$\Lambda$ and $\Sigma^0$ photoproduction from the proton.  It differs from 
traditional isobar approaches in its description of the non-resonant 
diagrams, which involve the exchange of $t$-channel $K$ and $K^*$ Regge 
trajectories.  A selection of $s$-channel resonances are then added to this 
background.  This ``Regge plus resonance'' approach has the advantage that 
the background diagrams contain only a few parameters that are constrained 
by high-energy data where the $t$-channel processes dominate.  In addition 
to the kaonic trajectories, this model includes the $s$-channel resonances 
$S_{11}$(1650), $P_{11}$(1710), $P_{13}$(1720), and $P_{13}$(1900).  Apart 
from these, the model includes either a $D_{13}$(1900) or $P_{11}$(1900) 
state in the $K^+\Lambda$ channel.

\boldmath
\section{CLAS $KY$ Electroproduction Results}
\unboldmath

CLAS has measured exclusive $K^+\Lambda$ and $K^+ \Sigma^0$ electroproduction 
on the proton for a range of momentum transfer $Q^2$ from 0.5 to 5.4~GeV$^2$ 
with electron beam energies from 2.6 to 5.7~GeV.  For this talk I will focus 
attention mainly on our 2.6~GeV data set. The final state hyperons were 
reconstructed from the $(e,e'K^+)$ missing mass, with an average hyperon 
resolution of $\sim$8~MeV.

The most general form for the virtual photoabsorption cross section of a kaon 
from an unpolarized proton target is given by:

\begin{displaymath}
\label{csec1}
\frac{d\sigma}{d\Omega_K^*} = \sigma_T + \epsilon \sigma_L
+ \epsilon \sigma_{TT} \cos 2\phi + \sqrt{\epsilon(1 + \epsilon)}
\sigma_{LT} \cos\phi + h \sqrt{\epsilon(1 - \epsilon)} \sigma_{LT'} 
\sin\phi.
\end{displaymath}

\noindent
In this expression, the cross section is decomposed into five structure
functions, $\sigma_T$, $\sigma_L$, $\sigma_{TT}$, $\sigma_{LT}$, and the
helicity-dependent $\sigma_{LT'}$ term, which are, in general, functions 
of $Q^2$, $W$, and $\theta_K^*$ only.  $\epsilon$ is the virtual photon 
polarization and $\Phi$ is the angle between the electron scattering and 
hadronic reaction planes.  One of the goals of the electroproduction 
program is to provide a detailed tomography of the structure functions vs. 
$Q^2$, $W$, and $\cos \theta_K^*$.  In the first phase of the analysis, we 
have measured the unseparated structure function ($\sigma_U = \sigma_T + 
\epsilon \sigma_L$) and, for the first time in the resonance region away from 
parallel kinematics, the interference structure functions $\sigma_{TT}$ and 
$\sigma_{LT}$.  Exploiting the $\Phi$ dependence of the reaction allows us to
extract the separate terms.  The $Q^2$ dependence of the data provides 
sensitivity to the associated form factors.  All of the published CLAS data 
are contained in the official CLAS data base\cite{database}.

\begin{figure}[htbp]
\vspace{10.9cm}
\includegraphics{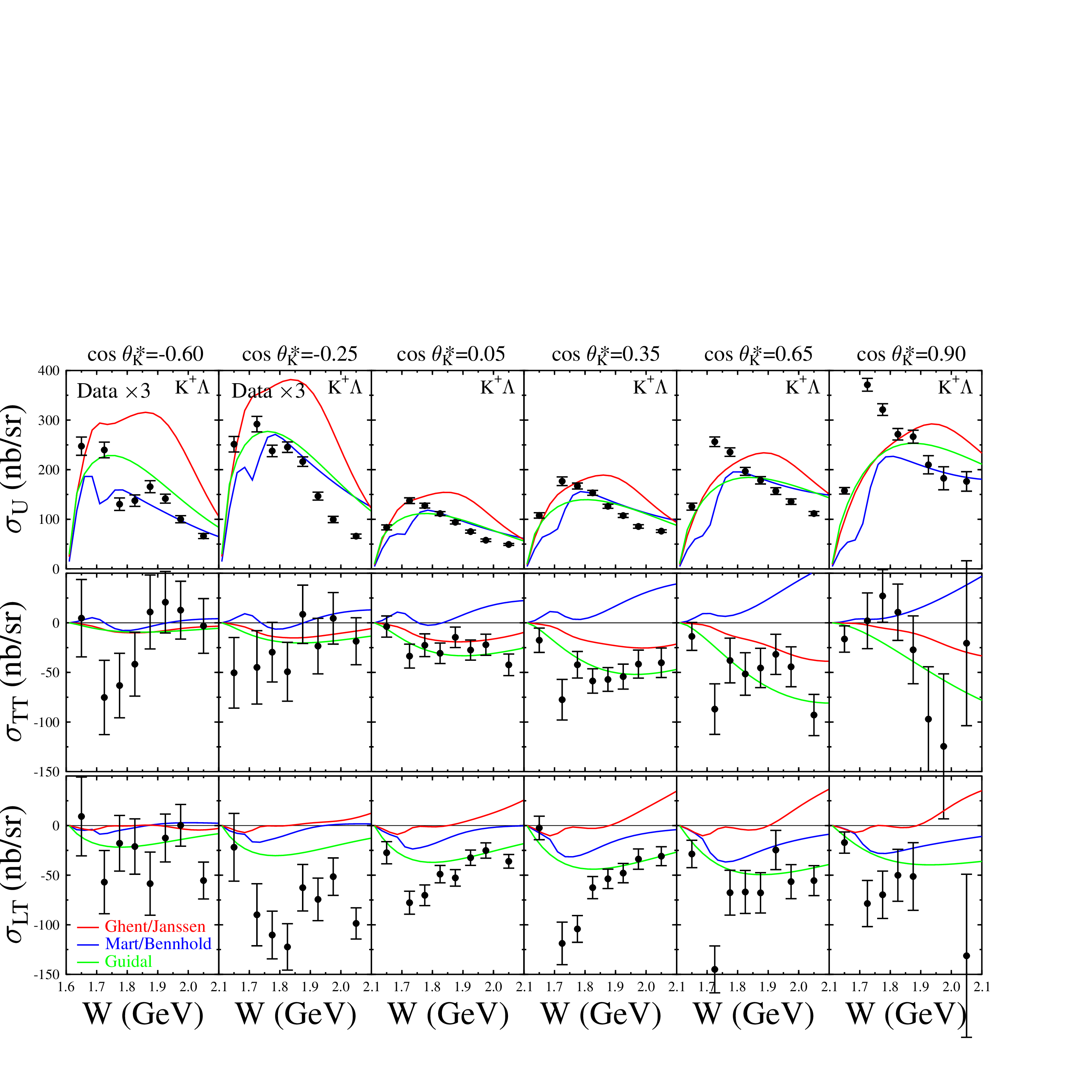}
\includegraphics{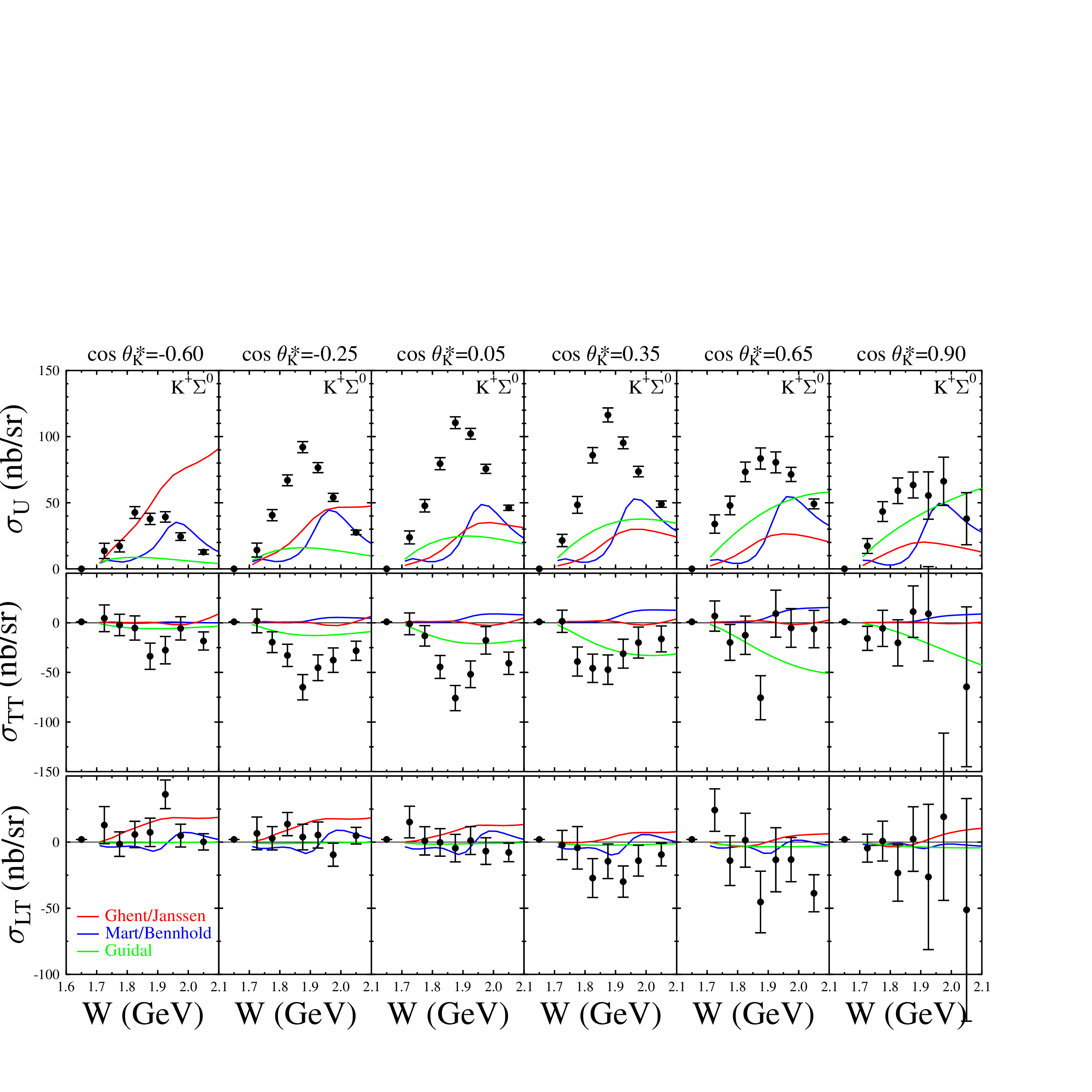}
\caption{Separated structure functions $\sigma_U$, $\sigma_{LT}$, and 
$\sigma_{TT}$ (nb/sr) vs. $W$ (GeV) for $K^+\Lambda$ (top) and $K^+\Sigma^0$ 
at 2.6~GeV and $Q^2$=0.65~GeV$^2$\cite{5st}.  The curves correspond to the 
indicated model calculations.}
\label{hyp_plot}
\end{figure}

A small sample of the available results from this analysis is shown in
Fig.~\ref{hyp_plot} vs. $W$ for each of our six angle bins for the kaon
\cite{5st}.  The kinematic dependence of the unpolarized structure functions 
shows that $\Lambda$ and $\Sigma^0$ hyperons are produced very differently.  
$\sigma_U$ at forward angles for $K^+\Lambda$ is dominated by a structure 
at $W$=1.7~GeV.  For larger kaon angles, a second structure emerges at about 
1.9~GeV, consistent with a similar signature in photoproduction.  
$\sigma_{TT}$ and $\sigma_{LT}$ are clearly non-zero and reflect the 
structures in $\sigma_U$.  The fact that $\sigma_{LT}$ is non-zero is 
indicative of longitudinal strength.  For the $K^+\Sigma^0$ final state, 
$\sigma_U$ is centrally peaked, with a single broad structure at 1.9~GeV.  
This is consistent with the photoproduction data.  $\sigma_{TT}$ reflects the 
features of $\sigma_U$, with $\sigma_{LT}$ consistent with zero everywhere, 
indicative of $\sigma_L$ being consistent with zero.

%%%%%%%%%%%%%%%%%%%%%%%%%%%%%%%%%%%%%%%%%%%%%%%%%%%%%%%%%%%%%%%%%%%%%%%%%%
%%%%%%%%%%%%%%%%%%%%%%%%%%%%%%%%%%%%%%%%%%%%%%%%%%%%%%%%%%%%%%%%%%%%%%%%%%

Using our data sets at 2.6 and 4.2~GeV, we have performed a Rosenbluth 
separation to extract $\sigma_L$ and $\sigma_T$ for several $W$ bins over 
the full kaon angular range for a single bin at $Q^2$=1.0~GeV$^2$ where the 
data sets overlap.  Our data indicate that $\sigma_L$ is consistent with zero
over all kinematics probed for the $K^+\Sigma^0$ final state.  For the
$K^+\Lambda$ final state, $\sigma_L$ is consistent with zero everywhere
except at our highest $W$ bin (1.95~GeV) and only a very forward kaon angles.
This analysis is consistent with our earlier results of $\sigma_L/\sigma_T$
extracted from our polarization data\cite{raue}.

%%%%%%%%%%%%%%%%%%%%%%%%%%%%%%%%%%%%%%%%%%%%%%%%%%%%%%%%%%%%%%%%%%%%%%%%%%
\begin{figure}[htbp]
\vspace{5.5cm}
\includegraphics{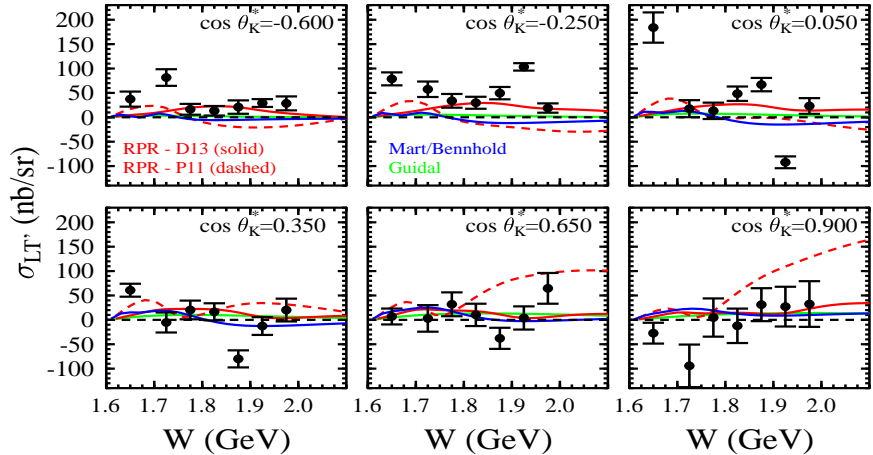}
\caption{Polarized structure function $\sigma_{LT'}$ (nb/sr) vs. $W$ (GeV)
for $K^+\Lambda$ for $Q^2$=1.0~GeV$^2$ and $\cos \theta_K^*$ points as 
indicated.  The curves correspond to the indicated model calculations.}
\label{fifth}
\end{figure}
%%%%%%%%%%%%%%%%%%%%%%%%%%%%%%%%%%%%%%%%%%%%%%%%%%%%%%%%%%%%%%%%%%%%%%%%%%

The polarized-beam asymmetry provides access to the fifth structure function 
$\sigma_{LT'}$. This observable probes imaginary parts of the interfering 
$L$ and $T$ amplitudes (as opposed to the real parts of the interference 
from $\sigma_{LT}$).  These imaginary parts vanish identically if the 
resonant state is determined by a single complex phase, which is the case for 
an isolated resonance.  A representative sample of our data at 2.6~GeV and 
$Q^2$=1.0~GeV$^2$ is shown in Fig.~\ref{fifth} for the $K^+\Lambda$ 
final state\cite{nasser}.  Note the strong interference effect seen at
central angles near 1.9~GeV.  The calculations shown are not able to
reproduce the features seen in the data.

\begin{figure}[htbp]
\vspace{5.6cm}
\includegraphics{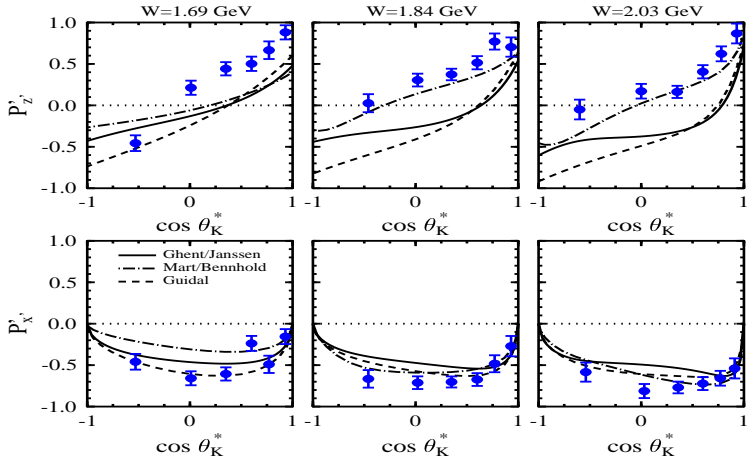}
\caption{CLAS transferred polarization\cite{carman03} from $\vec{e} p \to 
e'K^+ \vec{\Lambda}$ vs. $\cos \theta_K^*$ at 2.6~GeV for three different 
$W$ bins summed over $\Phi$ and $Q^2$.   The curves correspond to the 
indicated model calculations.}
\label{dpol}
\end{figure}

The first measurements of spin transfer from a longitudinally polarized 
electron beam to the $\Lambda$ hyperon produced in the exclusive 
$p(\vec{e},e'K^+)\vec{\Lambda}$ reaction have also been completed at 
CLAS\cite{carman03}.  A sample of the results highlighting the angular 
dependence of $P'$ summed over all $Q^2$ for three different $W$ bins is 
shown in Fig.~\ref{dpol} at 2.6~GeV.  The polarization along the virtual 
photon direction $P_{z'}'$ decreases with increasing $\theta_K^*$, while the 
orthogonal component in the hadronic reaction plane $P_{x'}'$ is constrained 
to be zero at $\cos \theta_K^*$ = $\pm$1 due to angular momentum conservation, 
and reaches a minimum at $\theta_K^* \sim 90^{\circ}$.  The component normal 
to the hadronic reaction plane $P_{y'}'$ is statistically consistent with 
zero as expected.  The accuracy of these data, coupled with the spread in the 
theory predictions, indicates that these data are sensitive to the 
resonant and non-resonant structure of the intermediate state.

%%%%%%%%%%%%%%%%%%%%%%%%%%%%%%%%%%%%%%%%%%%%%%%%%%%%%%%%%%%%%%%%%%%%%%%%%%
%%%%%%%%%%%%%%%%%%%%%%%%%%%%%%%%%%%%%%%%%%%%%%%%%%%%%%%%%%%%%%%%%%%%%%%%%%

\section{What Has Been Learned?}

According to the listings of the Particle Data Group\cite{pdg}, the
current knowledge of the $N^* \to KY$ and $\Delta^* \to K \Sigma$ 
couplings is quite limited.  Table~\ref{coup1} shows what has been 
determined experimentally.  The dominant couplings of $N^* \to K \Lambda$
include the $S_{11}$(1650), $P_{11}$(1710), and $P_{13}$(1720) resonances.
There are no known states that have been shown to couple to $K \Sigma$.
From the standpoint of $\Delta^* \to K \Sigma$ couplings, only the
$P_{33}$(1920) has a measured strength.  In addition, the available
photocoupling amplitudes that do exist, have rather large uncertainties.
Clearly there is significant room for improvement.

In the last several years, there has been some progress on the development
of $K^+\Lambda$ coupled-channels models based on fits to the available 
photoproduction data.  These include several thousand differential cross 
section and single and double polarization data points from CLAS, SAPHIR, 
and LEPS.  However, the lessons learned from these studies have not served 
to clarify our understanding of the $N^* \to K \Lambda$ couplings.  A 
coupled-channels model from Bonn\cite{sarantsev} has indicated that the most 
relevant states include the $P_{13}$(1720), $P_{11}$(1840), $D_{13}$(1870), 
$D_{13}$(2070), and $P_{13}$(2200).  A second coupled-channels model from 
Saghai {\it et al.}\cite{saghai_cc} has indicated that the most relevant 
states include the $S_{11}$(1535), $P_{13}$(1900), $D_{13}$(1520), 
$F_{13}$(1680), and $F_{17}$(1990), along with other required states given 
by $S_{11}$(1650), $F_{15}$(1680), $F_{15}$(2000), $D_{13}$(1954), 
$S_{11}$(1806), and $P_{13}$(1893).  The models have come to orthogonal 
conclusions as to the contributing states.  

%%%%%%%%%%%%%%%%%%%%%%%%%%%%%%%%%%%%%%%%%%%%%%%%%%%%%%%%%%%%%%%%%%%%%%%%%%%%%%%
\begin{table}[htpb]
\begin{center}
\begin{tabular} {|c|c|c|c|} \hline
State                  & PDG  & B.R. ($K\Lambda$) & B.R. ($K\Sigma$) \\ \hline
$N^*(1650)$ $S_{11}$      & **** & 3-11\% & - \\ \hline
$N^*(1675)$ $D_{15}$      & **** & $<$1\% & - \\ \hline
$N^*(1680)$ $F_{15}$      & **** & -      & - \\ \hline
$N^*(1700)$ $D_{13}$      & ***  & $<$3\% & - \\ \hline
$N^*(1710)$ $P_{11}$      & ***  & 5-25\% & - \\ \hline
$N^*(1720)$ $P_{13}$      & ***  & 1-15\% & - \\ \hline
$N^*(1900)$ $P_{13}$      & **   & -      & - \\ \hline
$N^*(1990)$ $F_{17}$      & **   & -      & - \\ \hline
$N^*(2000)$ $F_{15}$      & **   & -      & - \\ \hline \hline
$\Delta^*(1900)$ $S_{31}$ & **   &        & -     \\ \hline
$\Delta^*(1905)$ $F_{35}$ & **** &        & -     \\ \hline
$\Delta^*(1910)$ $P_{31}$ & **** &        & -     \\ \hline
$\Delta^*(1920)$ $P_{33}$ & ***  &        & 2.1\% \\ \hline
$\Delta^*(1930)$ $D_{35}$ & ***  &        & -     \\ \hline
$\Delta^*(1940)$ $D_{33}$ & *    &        & -     \\ \hline
$\Delta^*(1950)$ $F_{37}$ & **** &        & -     \\ \hline
\end{tabular}			  
\end{center}			  
\caption{$N^*$ and $\Delta^*$ states and their known branching ratios 
into $K\Lambda$ and $K\Sigma$ final states\cite{pdg}.  The second column gives
the PDG star-rating for the states.}
\label{coup1}
\end{table}
%%%%%%%%%%%%%%%%%%%%%%%%%%%%%%%%%%%%%%%%%%%%%%%%%%%%%%%%%%%%%%%%%%%%%%%%%%%%%%%

Finally, a new multipole model from Mart and Sulaksono\cite{multipole} has 
determined that the extracted $N^* \to K\Lambda$ couplings are highly 
dependent on the data used as input.  Based on fits to the SAPHIR and LEPS 
photoproduction data, they have found that the most relevant states include 
the $S_{11}$(1650), $P_{13}$(1720), $D_{13}$(1700), $D_{13}$(2080), 
$F_{15}$(1680), and $F_{15}$(2000).  However, using the CLAS and LEPS 
photoproduction data, they find instead that the dominant states include the 
$P_{13}$(1900), $D_{13}$(2080), $D_{15}$(1675), $F_{15}$(1680), and 
$F_{17}$(1990).  The difficulties here arise due to important shape 
differences between the CLAS and SAPHIR angular distributions.  CLAS is
continuing to examine this issue.

Ultimately, the $KY$ photo- and electroproduction data should be fit 
simultaneously with a full set of channels (e.g. $\pi N$, $\eta N$, 
$\omega N$, $\phi N$, $\pi \pi N$).  The inclusion of electroproduction 
data should provide powerful new information to better separate the resonant 
and non-resonant contributions.  At the current time, this approach is too 
difficult.  However, several groups are working in this direction, and we 
expect that a more complete coupled-channels approach will be possible in 
the not too distant future.

\section{Future Plans}

Studies of strangeness final states represent an important part of the 
overall CLAS physics program.  A variety of observables have been published 
to date for both our photo- and electroproduction data.  In the near future, 
CLAS will provide additional new high quality data in the $W$ range from 1.6 
to 2.4~GeV using both circularly and linearly polarized photons.  Three 
measurement programs are currently in progress to study strangeness physics 
with CLAS.

The first program is called g13.  This experiment, which completed a long
run with CLAS in the first half of 2007, was designed to study $\gamma n$
interactions on an unpolarized deuterium target.  This experiment will study 
$K^0 \Lambda$, $K^0 \Sigma^0$, and $K^+ \Sigma^-$ final states.  The second 
program is called g9, and will take data in CLAS in the second half of 2007.  
This experiment will study $K^+\Lambda$ and $K^+ \Sigma^0$ final states.  
The target for this experiment employs a novel frozen spin target that allows 
studies with both longitudinally and transversely polarized protons.  The 
final program will employ the HD-ice target from Brookhaven National Lab
\cite{hdice} to study $\gamma n \to K^0 \Lambda$, $K^0 \Sigma^0$, and 
$K^+ \Sigma^-$ reactions.  This target is now in the process of being 
modified for use in CLAS.

The combined FROST and HD-ice programs will provide a complete set of 
observables with high statistics in both the $\gamma p$ and $\gamma n$ 
channels for both the $K \Lambda$ and $K \Sigma$ final states.  The ability 
to take data for all combinations of beam, target, and recoil polarization 
observables with the same detector will allow systematics to be minimized.
These high profile programs at JLab will provide important input to 
disentangle the $N^*$ spectrum.

At the present time, JLab is well underway with its plans to upgrade the
accelerator from a maximum energy of 6~GeV to a maximum energy of 12~GeV.  
Along with the accelerator upgrade, the experimental halls will also be 
upgraded.  The CLAS detector will be modified to a new configuration called 
CLAS12.  The new large acceptance detector will still be based on a toroidal
design, but the new magnet will be outfitted with new drift chambers and 
a new central detector system.  The calorimetry, time-of-flight, and
{\v C}erenkov detector systems of CLAS will also undergo significant
upgrades.

The new CLAS12 experiment is presently slated to begin its physics
program in 2014.  An important aspect of the program is the
continued study of strangeness physics.  This program includes both
semi-inclusive and exclusive measurements focussing on spectroscopy,
quark distribution functions, and deep-inelastic scattering.

\section{Summary and Conclusions}

In this talk I have reviewed some of the key reasons why the photo-
and electroproduction processes of open-strangeness production are 
important for the investigation of baryonic structure and missing 
quark model states.  I have discussed several aspects of the 
CLAS strangeness physics program highlighting the breadth and quality 
of our photo- and electroproduction data sets on the nucleon.  These data
will provide not only high statistics differential cross sections, but
high precision data for all combinations of beam, target, and recoil
polarization observables on the proton and neutron.  Our analyses indicate 
that the data are highly sensitive to the ingredients of the models, including 
the specific baryonic resonances included, along with their associated 
form factors and coupling constants.  New complete amplitude-level analyses 
are called for to more fully unravel the contributions to the intermediate 
state.

This work has been supported by the U.S. Department of Energy and the 
National Science Foundation.

\end{document}